\pdfoutput=1
\documentclass[a4paper]{jpconf}
\usepackage{graphicx}
\usepackage{amssymb,amsmath,cite}

\newcommand{\ba}{\begin{eqnarray}}
\newcommand{\ea}{\end{eqnarray}}
\newcommand{\ban}{\begin{eqnarray*}}
\newcommand{\ean}{\end{eqnarray*}}
\newcommand{\bsub}{\begin{subequations}}
\newcommand{\esub}{\end{subequations}}

\def\ket#1{|#1\rangle}
\def\bra#1{\langle#1|}
\def\lm{(\lambda,\mu)}
\def\Htwo{\hat{H}_2(\xi)}

\begin{document}
\title{Order, chaos and quasi symmetries in \\
a first-order quantum phase transition}

\author{A Leviatan and M Macek}

\address{Racah Institute of Physics, The Hebrew University, 
Jerusalem 91904, Israel}

\ead{ami@phys.huji.ac.il, mmacek@phys.huji.ac.il}

\begin{abstract}
We study the competing order and chaos in a 
first-order quantum phase transition with a high barrier. 
The boson model Hamiltonian employed, interpolates 
between its U(5) (spherical) and SU(3) (deformed) limits.
A classical analysis reveals regular (chaotic) dynamics at low (higher) 
energy in the spherical region, coexisting with a robustly regular dynamics 
in the deformed region. A quantum analysis discloses, amidst a 
complicated environment, persisting regular multiplets of states 
associated with partial U(5) and quasi SU(3) dynamical symmetries.
\end{abstract}
\section{Introduction}
\label{sec:intro}

Quantum phase transitions (QPTs) are qualitative changes in the 
ground state properties of a physical system induced by a variation 
of parameters $\lambda$ in the 
quantum Hamiltonian $\hat{H}(\lambda)$~\cite{Hert76,Gilm79}. 
Such ground-state transformations have received considerable 
attention in recent years and have found a variety of applications in 
many areas of physics and chemistry~\cite{carr}. 
The particular type of QPT is reflected in the topology of the 
underlying mean-field (Landau) potential $V(\lambda)$. 
Most studies have focused on second-order 
(continuous) QPTs~\cite{Sachdev99}, 
where $V(\lambda)$ has a single minimum 
which evolves continuously into another minimum. 
The situation is more complex for 
discontinuous (first-order) QPTs, where 
$V(\lambda)$ develops multiple minima 
that coexist in a range of $\lambda$ values 
and cross at the critical point, $\lambda\!=\!\lambda_c$.
The interest in such QPTs stems from their key role in 
phase-coexistence phenomena at zero temperature.
Examples are offered by the metal-insulator Mott 
transition~\cite{maria04}, heavy-fermion superconductors~\cite{Pf09}, 
quantum Hall bilayers~\cite{karm09}
and shape-coexistence in mesoscopic systems, 
such as atomic nuclei~\cite{Cejn10}.

The competing interactions in the Hamiltonian that drive these 
ground-state phase transitions can affect dramatically the nature of 
the dynamics and, in some cases, lead to the emergence of quantum chaos. 
This effect has been observed in quantum optics models of $N$ 
two-level atoms interacting with a 
single-mode radiation field~\cite{Emar03}, where 
the onset of chaos is triggered by continuous QPTs.
In the present contribution, we address the mixed regular and chaotic 
dynamics associated with a first order 
QPT~\cite{MacLev11,LevMac12,MacLev13}.
For that purpose, we employ an interacting boson model 
which describes such QPTs between spherical and axially-deformed nuclei.
The model is amenable to both classical and quantum treatments, has 
a rich group structure and inherent geometry, which makes it an ideal 
framework for studying the intricate interplay of order and chaos and 
the role of symmetries in such shape-phase transitions.  

\section{Symmetries, geometry and quantum phase transitions in the IBM}

The interacting boson model (IBM)~\cite{ibm} describes low-lying 
quadrupole collective states in nuclei in terms of $N$ interacting monopole 
$(s)$ and quadrupole $(d)$ bosons representing valence nucleon pairs.
The bilinear combinations 
${\cal G}_{ij}\equiv b^{\dag}_{i}b_j =
\{s^{\dag}s,\,s^{\dag}d_{m},\, d^{\dag}_{m}s,\, 
d^{\dag}_{m}d_{m '}\}$ span a~U(6) algebra, which 
serves as the spectrum generating algebra. 
The IBM Hamiltonian is expanded in terms of these generators, 
$\hat{H} = \sum_{ij}\epsilon_{ij}\,{\cal G}_{ij} 
+ \sum_{ijk\ell}u_{ijk\ell}\,{\cal G}_{ij}{\cal G}_{k\ell}$, 
and consists of Hermitian, 
rotational-invariant interactions 
which conserve the total number of $s$- and $d$- bosons, 
$\hat N = \hat{n}_s + \hat{n}_d = 
s^{\dagger}s + \sum_{m}d^{\dagger}_{m}d_{m}$. 
A dynamical symmetry (DS) occurs if the Hamiltonian
can be written in terms of the Casimir operators 
of a chain of nested sub-algebras of U(6).
The Hamiltonian is then completely solvable in the basis associated with 
each chain. 
The three dynamical symmetries of the IBM and corresponding bases are
\bsub
\label{eq:chains}
\ba
&&{\rm U(6)} \supset {\rm U(5)}  \supset {\rm O(5)} \supset {\rm O(3)}
\qquad
\vert N,n_d,\tau,n_{\Delta},L\rangle 
\;\;\;\;\,
\label{u5ds} 
\;{\rm spherical\;vibrator}
\\
&&{\rm U(6)} \supset {\rm SU(3)} \supset {\rm O(3)}
\qquad\qquad\;\;\, 
\vert N,(\lambda,\mu),K,L\rangle \quad\;\;
\label{su3ds} 
{\rm axially} {\rm-deformed \; rotor}
\\
&&{\rm U(6)} \supset {\rm O(6)}  \supset {\rm O(5)} \supset {\rm O(3)}
\qquad
\vert N,\sigma,\tau,n_{\Delta},L\rangle \qquad
\,\gamma{\rm-unstable\; deformed\; rotor}\qquad\quad
\label{o6ds}
\ea
\esub
The associated analytic solutions 
resemble known limits of the geometric model of nuclei~\cite{bohr98}, 
as indicated above. The basis members are 
classified by the irreducible representations (irreps) of the 
corresponding algebras. 
Specifically, 
the quantum numbers $N,n_d,(\lambda,\mu),\sigma,\tau$ and $L$ 
label the relevant irreps of ${\rm U(6),U(5),SU(3),O(6),O(5)}$ 
and ${\rm O(3)}$, 
respectively. 
$n_{\Delta}$ and $K$ are multiplicity labels needed for complete 
classification in the reductions ${\rm O(5)}\supset {\rm O(3)}$ and 
${\rm SU(3)}\supset {\rm O(3)}$, respectively.
Each basis is complete and can be used for a numerical diagonalization 
of the Hamiltonian in the general case.
A geometric visualization of the model is obtained by 
a potential surface, 
$V(\beta,\gamma)\!=\! \bra{\beta,\gamma;N}\hat{H}\ket{\beta,\gamma;N}$, 
defined by the expectation value of the Hamiltonian in the 
intrinsic condensate state~\cite{Gino80,Diep80}
\bsub
\label{condgen}
\ba
\vert\beta,\gamma ; N \rangle &=& 
(N!)^{-1/2}[\,\Gamma^{\dagger}_{c}(\beta,\gamma)\,]^N\vert 0\rangle ~,\\
\Gamma^{\dagger}_{c}(\beta,\gamma) &=& 
\left [\beta\cos\gamma d^{\dagger}_{0} + \beta\sin{\gamma} 
( d^{\dagger}_{2} + d^{\dagger}_{-2})/\sqrt{2} 
+ \sqrt{2-\beta^2}s^{\dagger}\right ]/\sqrt{2} ~.
\ea
\esub
Here $(\beta,\gamma)$ are quadrupole shape parameters 
analogous to the variables of the collective model of nuclei.
Their values 
$(\beta_{\mathrm{eq}},\gamma_{\mathrm{eq}})$ at the global minimum 
of $V(\beta,\gamma)$ 
define the equilibrium shape for a given Hamiltonian.
For one- and two-body interactions, 
the shape can be spherical $(\beta_{\mathrm{eq}}=0)$ or 
deformed $(\beta_{\mathrm{eq}}>0)$ with 
$\gamma_{\mathrm{eq}}=0$ (prolate), 
$\gamma_{\mathrm{eq}}=\pi/3$ (oblate), 
or $\gamma$-independent. 

The dynamical symmetries of 
Eq.~(\ref{eq:chains}), correspond to phases of the system, and provide 
analytic benchmarks for the dynamics of stable nuclear shapes. 
Quantum phase transitions (QPTs) between 
such stable shapes have been studied extensively
in the IBM framework~\cite{Diep80,Cejnar09}
and are manifested empirically in nuclei~\cite{Cejn10}.
The Hamiltonians employed mix interaction terms from 
different DS chains, {\it e.g.}, 
$\hat{H}(\lambda) = \lambda\hat{H}_a + (1-\lambda)\hat{H}_b$. 
The coupling coefficient $(\lambda$) responsible 
for the mixing, serves as the control parameter 
and the surface, $V(\lambda)\equiv V(\lambda;\beta,\gamma)$, serves as 
the Landau potential. 
In general, under such circumstances, solvability is lost, there 
are no remaining non-trivial conserved quantum numbers 
and all eigenstates are expected to be mixed.
However, for particular symmetry breaking, some intermediate symmetry 
structure can 
survive. The latter include partial dynamical symmetry (PDS)~\cite{Lev11} 
and quasi-dynamical symmetry (QDS)~\cite{Rowe04}. 
In a PDS, the conditions of an 
exact dynamical symmetry (solvability of the complete spectrum and 
existence of exact quantum numbers for all eigenstates) are relaxed and 
apply to only part of the eigenstates. 
In a QDS, particular states continue to exhibit 
selected characteristic properties ({\it e.g.}, energy 
and B(E2) ratios) of the closest dynamical symmetry, in the face 
of strong-symmetry breaking interactions.
This ``apparent'' symmetry is due to the coherent nature of the mixing.
As discussed below, 
both PDS and QDS are relevant to 
quantum phase transitions~\cite{Rowe04,Lev07}. 

In view of the central role of the Landau potential, $V(\beta,\gamma)$, 
for QPTs, it is convenient to resolve the Hamiltonian into two 
parts,
$\hat{H} = \hat{H}_{\mathrm{int}} + \hat{H}_{\mathrm{col}}$~\cite{Lev87}. 
The intrinsic part ($\hat{H}_{\mathrm{int}}$) 
determines the potential surface 
while the collective part ($\hat{H}_{\mathrm{col}}$)
is composed of kinetic terms which do not affect the shape of 
$V(\beta,\gamma)$. For first-order QPTs, the resolution allows the 
construction of an intrinsic Hamiltonian with a high-barrier~\cite{Lev06}, 
and by treating it separately, one avoids the complication of 
rotation-vibration couplings that can obscure the simple pattern 
of mixed dynamics, reported below. 
Henceforth, we confine the discussion to the dynamics of the 
intrinsic Hamiltonian.

\section{Intrinsic Hamiltonian for a first-order QPT}
\label{sec:QPT1}

Focusing on first-order QPTs between 
stable spherical ($\beta_{\mathrm{eq}}=0$) and prolate-deformed 
($\beta_{\mathrm{eq}}>0$, $\gamma_{\mathrm{eq}}=0$) shapes, 
the intrinsic Hamiltonian reads
\bsub
\label{eq:Hint}
\ba
\hat{H}_1(\rho)/\bar{h}_2 &=& 
2(1\!-\! 2\rho^2)\hat{n}_d(\hat{n}_d \!-\! 1)
+ 2 R^{\dag}_2(\rho) \cdot\tilde{R}_2(\rho) ~,
\label{eq:H1} \\
\hat{H}_2(\xi)/ \bar{h}_2 &=& 
\xi P^{\dag}_0 P_0 + 
P^{\dag}_2 \cdot \tilde{P}_2 ~,
\label{eq:H2} 
\ea
\esub
where $\hat{n}_d$ 
is the $d$-boson number operator, 
$R^{\dag}_{2\mu}(\rho) \!=\! \sqrt{2}s^\dag d^\dag_\mu + 
\rho\sqrt{7}(d^\dag d^\dag)^{(2)}_\mu$, 
$P^{\dag}_0 \!=\! d^\dag \cdot d^\dag - 2 (s^\dag)^2$ 
and 
$P^{\dag}_{2\mu} \!=\! 2 s^\dag d^\dag_\mu + 
\sqrt{7}(d^\dag d^\dag)^{(2)}_\mu$.
Here
$\tilde{R}_{2\mu} \!=\! (-1)^{\mu}R_{2,-\mu}$, 
$\tilde{P}_{2\mu} \!=\! (-1)^{\mu}P_{2,-\mu}$ 
and the dot implies a scalar product. 
Scaling by $\overline{h}_2\equiv h_2/ N(N-1)$ 
is used throughout, to facilitate the comparison with the classical limit. 
The control parameters that drive the QPT are $\rho$ and $\xi$, 
with $0\leq\rho\leq\tfrac{1}{\sqrt{2}}$ and $0\leq\xi\leq 1$.

The intrinsic Hamiltonian in the spherical 
phase, $\hat{H}_{1}(\rho)$, has by construction 
the intrinsic state of Eq.~(\ref{condgen}) 
with $\beta_{\mathrm{eq}}\!=\!0$ as zero energy eigenstate.
For large $N$, its normal modes involve quadrupole vibrations
about the spherical global minimum of the potential surface, 
with frequency $\epsilon \!=\! 4\bar{h}_2N$. 
For $\rho=0$ it reduces to 
\ba
\hat{H}_1(\rho=0)/\bar{h}_2 &=&
2\hat{n}_d(\hat{n}_d - 1)
+4 (\hat{N}-\hat{n}_d)\hat{n}_d ~.
\label{H1u5}
\ea
Since $\hat{n}_d$ is the linear Casimir operator of U(5), 
$\hat{H}_1(\rho=0)$ has U(5) dynamical symmetry (DS). 
The spectrum is completely solvable
$E_{\rm DS} = [2n_d(n_d-1) + 4(N-n_d)n_d]\bar{h}_2$, 
and the eigenstates, $\ket{N,n_d,\tau,n_{\Delta},L}$, 
are those of the U(5) chain, Eq.~(\ref{u5ds}). 
The spectrum resembles that of an anharmonic spherical vibrator, 
describing quadrupole excitations of a spherical equilibrium shape.
The lowest U(5) multiplets involve states with quantum numbers 
$(n_d\!=\!0,\tau\!=\!0,L\!=\!0)$, $(n_d\!=\!1,\tau\!=\!1,L\!=\!2)$, 
$(n_d\!=\!2,\tau\!=\!2,L\!=\!2,4;\tau=0,L=0)$, 
$(n_d\!=\!3,\tau\!=\!3,L\!=\!6,4,3,0;\tau\!=\!1,L\!=\!2)$.

For $\rho > 0$, $\hat{H}_{1}(\rho)$ has an additional 
$\rho [(d^{\dag}d^{\dag})^{(2)}\cdot \tilde{d}s + 
s^{\dag}d^{\dag}\cdot(\tilde{d}\tilde{d})^{(2)}]$ term, 
which breaks the U(5) DS, 
and induces U(5) and O(5) mixing subject to
$\Delta n_d = \pm 1$ and 
$\Delta\tau=\pm 1, \pm 3$. The explicit breaking of O(5) symmetry 
leads to non-integrability and, as will be shown in subsequent discussions, 
is the main cause for the onset of chaos in the spherical region.
Although $\hat{H}_1(\rho>0)$ is not diagonal in the 
U(5) chain, it retains the following selected solvable U(5) basis states 
\bsub
\ba
\label{ePDSu5L0}
\vert N, n_d=\tau=L=0\rangle \;\;\;\; 
&&E_{\rm PDS} = 0 ~,\\
\vert N, n_d=\tau=L=3\rangle \;\;\;\; 
&&E_{\rm PDS} 
= 12\left ( N -2 + 3\rho^2 \right )\bar{h}_2 ~,
\label{ePDSu5L3}
\ea
\label{ePDSu5}
\esub
while other eigenstates are mixed with respect to U(5). 
As such, it exhibits U(5) partial dynamical symmetry [U(5)-PDS].

The intrinsic Hamiltonian in the deformed 
phase, $\hat{H}_{2}(\xi)$, has by construction 
the intrinsic state of Eq.~(\ref{condgen}) 
$\vert\beta_{\mathrm{eq}}=\tfrac{2}{\sqrt{3}},\gamma_{\mathrm{eq}}=0 ; N \rangle$
as zero energy eigenstate. 
For large $N$, its normal modes involve 
both $\beta$ and $\gamma$ vibrations 
about the deformed  global minimum of $V(\beta,\gamma)$, 
with frequencies $\epsilon_\beta \!=\! 4\bar{h}_2 N(2\xi + 1)$ and 
$\epsilon_\gamma \!=\! 12 \bar{h}_2 N$. 
For $\xi=1$, the Hamiltonian reduces to
\ba
\hat{H}_2(\xi=1) /\bar{h}_2 &=& 
-\hat{C}_{\rm SU(3)} + 2\hat{N}(2\hat{N}+3) ~,
\ea
It involves the quadratic Casimir of SU(3) and hence has SU(3) DS. 
The spectrum is completely solvable, 
$E_{\rm DS} /\bar{h}_2 = [-(\lambda^2 + \mu^2+\lambda\mu + 3\lambda+3\mu)
+ 2N(2N+3)]\bar{h}_2$, 
and the eigenstates, $\ket{N,(\lambda,\mu),K,L}$, 
are those of the SU(3) chain~, Eq.~(\ref{su3ds}). 
The spectrum resembles that of an axially-deformed rotor with 
degenerate $K$-bands arranged in SU(3) $(\lambda,\mu)$ multiplets, 
$K$ being the angular momentum projection on the symmetry axis. 
The rotational states in each band have angular momenta 
$L=0,\, 2,\,4\ldots$, for $K=0$ and 
$L=K,K+1,K+2,\ldots$, for $K>0$.
The lowest SU(3) multiplets are $(2N,0)$ 
which contains the ground band $g(K=0)$, 
and $(2N-4,2)$ which contains the $\beta(K=0)$ and $\gamma(K=2)$ 
bands.

For $\xi < 1$, $\hat{H}_2(\xi)$ has an additional term,
$(\xi-1) P^{\dag}_0P_0$, which breaks the SU(3) DS 
and most eigenstates are mixed with respect to SU(3). However, 
the following states 
\bsub
\ba
&&\vert N,(2N,0)K=0,L\rangle \;\;\;\; 
L=0,2,4,\ldots, 2N
\nonumber\\
&&E_{\rm PDS} = 0
\label{solsu3g}
\\
&&\vert N,(2N-4k,2k)K=2k,L\rangle
\;\;\;\;
L=K,K+1,\ldots, (2N-2k) \qquad\; k>0 ~,
\qquad\qquad
\nonumber\\
&&
E_{\rm PDS} =  6k \left (2N - 2k+1 \right )\bar{h}_2
\label{solsu3gam}
\ea
\label{solsu3}
\esub
remain solvable with good SU3) symmetry. As such, 
$\hat{H}_2(\xi<1)$ exhibits SU(3) partial dynamical 
symmetry [SU(3)-PDS]. The selected states of Eq.~(\ref{solsu3}) 
span the ground band $g(K=0)$ and $\gamma^{k}(K=2k)$ bands.

The intrinsic Hamiltonians, $\hat{H}_{1}(\rho)$ and $\hat{H}_{2}(\xi)$ 
of Eq.~(\ref{eq:Hint}), 
with $0\leq\rho\leq\tfrac{1}{\sqrt{2}}$ and $0\leq\xi\leq 1$, 
interpolate between the U(5) $(\rho\!=\!0)$ and 
SU(3) $(\xi\!=\!1)$ DS limits. 
The two Hamiltonians coincide at the critical point 
$\rho_c\!=\!\tfrac{1}{\sqrt{2}}$ and $\xi_c \!=\!0$: 
$\hat{H}_{1}(\rho_c) = \hat{H}_{2}(\xi_c)$. 
Both Hamiltonians support subsets of solvable 
PDS states, Eqs.~(\ref{ePDSu5}) and (\ref{solsu3}), 
whose analytic properties provide unique signatures for their 
identification in the quantum spectrum.

\section{Classical limit and topology of the Landau potential}
\label{sec:Class}

The classical limit of the IBM is obtained through the use of Glauber 
coherent states. This amounts to 
replacing $(s^{\dagger},\,d^{\dagger}_{\mu})$ by six 
c-numbers $(\alpha_{s}^{*},\,\alpha_{\mu}^{*})$ rescaled 
by $\sqrt{N}$ and taking $N\rightarrow\infty$, with $1/N$ playing the 
role of $\hbar$~\cite{Hatch82}. Number conservation ensures that 
phase space is 10-dimensional and can be phrased in terms of 
two shape (deformation) variables, three orientation (Euler) angles 
and their conjugate momenta. The shape variables can be identified with the 
$\beta, \gamma$ variables introduced through Eq.~(\ref{condgen}). 
Setting all momenta to zero, yields the classical potential which 
is identical to $V(\beta,\gamma)$ mentioned above. 
In the classical analysis presented below we consider, for simplicity, the 
dynamics of $L=0$ vibrations, for which only two 
degrees of freedom are active. 
The rotational dynamics with $L>0$ is examined in 
the subsequent quantum analysis. 

For the intrinsic Hamiltonian of Eq.~(\ref{eq:Hint}), constrained to $L=0$,  
the above procedure yields the following classical Hamiltonian
\bsub
\label{eq:Hintcl}
\ba
\label{eq:H1cl}
\mathcal{H}_1 (\rho)/h_2 &=& 
{\cal H}_{d,0}^2 + 2(1 - {\cal H}_{d,0}){\cal H}_{d,0} 
+ 2 \rho^2  p_\gamma^2
\nonumber\\
&& + \rho\sqrt{2(1 - {\cal H}_{d,0})}
\left [\,(p_\gamma^2/\beta - \beta p^2_{\beta} - \beta^3) \cos{3\gamma} 
+ 2p_{\beta} p_\gamma \sin{3\gamma}\,\right] ~,\\
\label{eq:H2cl}
\mathcal{H}_2(\xi)/h_2 &=& {\cal H}_{d,0}^2 
+ 2(1\!\!-\!\!{\cal H}_{d,0}){\cal H}_{d,0} + p_\gamma^2 
\nonumber\\
&&+ \sqrt{1 - {\cal H}_{d,0}}
\left [\,(p_\gamma^2/\beta 
- \beta p^2_{\beta} - \beta^3) \cos{3\gamma} 
+ 2p_{\beta} p_\gamma \sin{3\gamma}\,\right ]\qquad  
\nonumber \\
&& + \xi\left [\,\beta^2 p_\beta^2 + \tfrac{1}{4}(\beta^2 - T)^2 
- 2(1 - {\cal H}_{d,0})(\beta^2 - T) 
+ 4(1\!\!-\!\!{\cal H}_{d,0})^2\,\right] ~.
\ea
\esub
Here the coordinates $\beta\in[0,\sqrt{2}]$, $\gamma\in[0,2\pi)$ 
and their canonically 
conjugate momenta $p_\beta\in[0,\sqrt{2}]$ and $p_\gamma\in[0,1]$ 
span a compact classical phase space. 
The term, ${\cal H}_{d,0} = (T+\beta^2)/2$ with 
$T = p_{\beta}^2+p_\gamma^2/\beta^2$, 
denotes the classical limit of $\hat{n}_d$ 
(restricted to $L=0$) and forms an isotropic harmonic oscillator 
Hamiltonian in the $\beta$ and $\gamma$ variables. 
Notice that the classical Hamiltonian of Eq.~(\ref{eq:Hintcl}) contains
complicated momentum-dependent terms originating from the two-body 
interactions in the Hamiltonian~(\ref{eq:Hint}), not just the usual 
quadratic kinetic energy $T$. 
Setting $p_{\beta} = p_\gamma=0$ in Eq.~(\ref{eq:Hintcl}) leads to the 
following classical potential
\bsub
\label{eq:Vcl}
\ba
\label{eq:V1}
V_1(\rho)/h_2 &=& 
2 \beta^2 - 2\rho \sqrt{2\!-\!\beta^2} \beta^3\cos3\gamma 
- \textstyle{\frac{1}{2}}\beta^4 ~,\\
\label{eq:V2}
V_2(\xi)/h_2 &=&  
2(1 - 3\xi) \beta^2 
-\sqrt{2(2\!-\!\beta^2)} \beta^3\cos3\gamma 
+ \textstyle{\frac{1}{4}} (9 \xi - 2)\beta^4 + 4\xi ~.
\ea
\esub
The variables $\beta$ and $\gamma$ 
can be interpreted as polar coordinates 
in an abstract plane parametrized by Cartesian coordinates 
$x\!=\!\beta\cos{\gamma}$ and $y\!=\!\beta\sin{\gamma}$.
Using these relations together with 
$p_x \!=\! p_{\beta}\cos\gamma  - (p_{\gamma}/\beta)\sin\gamma$
and $p_y \!=\! (p_{\gamma}/\beta)\cos\gamma + p_{\beta}\sin\gamma$, 
one can cast the classical Hamiltonian, Eq.~(\ref{eq:Hintcl}) and 
potential, Eq.~(\ref{eq:Vcl}), in Cartesian form. Thus, 
${\cal H}_{d,0} = (p_{x}^2 + p_{y}^2 + x^2 + y^2)/2$ and 
the potentials $V(\beta,\gamma)=V(x,y)$ depend on the combinations
$ \beta^2 \!=\! x^2 + y^2$, $\beta^4= (x^2+y^2)^2$ and 
$\beta^3\cos 3\gamma \!=\! x^3 - 3xy^2$.
 
The control parameters $\rho$ and $\xi$ determine the 
landscape and extremal points of the potentials 
$V_{1}(\rho;\beta,\gamma)$ and $V_{2}(\xi;\beta,\gamma)$, 
Eq.~(\ref{eq:Vcl}). 
Important values of these parameters at which 
a pronounced change in structure is observed, 
are the spinodal point $(\rho^{*})$ where a second (deformed) 
minimum occurs, an anti-spinodal point $(\xi^{**})$ 
where the first (spherical) minimum 
disappears and a critical point $(\rho_c,\xi_c)$ 
in-between, where the two minima are degenerate. 
For the potentials under discussion, the values of the control 
parameters at these points are
\ba
\rho^{*} = \tfrac{1}{2}\;\;\; , \;\;\; 
(\rho_c=\tfrac{1}{\sqrt{2}}\,,\,\xi_c=0) \;\;\; , \;\;\;
\xi^{**} = \tfrac{1}{3} \;\;\; .
\label{points}
\ea 
The critical point separates the spherical and deformed phases. 
The spinodal and anti-spinodal points embrace it and mark the 
boundary of the phase coexistence region. 

In general, the only $\gamma$-dependence in the potentials~(\ref{eq:Vcl})
is due to the $\sqrt{2-\beta^2}\beta^3\cos 3\gamma$ term. 
This induces a three-fold symmetry about the origin $\beta\!=\!0$. 
As a consequence, the deformed extremal points are obtained for 
$\gamma\!=\!0,\tfrac{2\pi}{3},\tfrac{4\pi}{3}$ (prolate shapes), 
or $\gamma \!=\! \tfrac{\pi}{3},\textstyle{\pi},\tfrac{5\pi}{3}$ 
(oblate shapes).
It is therefore possible to restrict the 
analysis to $\gamma=0$ and allow for   
both positive and negative values of $\beta$, corresponding to prolate 
and oblate deformations, respectively. 
The potentials $V(\beta,\gamma=0)\!=\!V(x,y=0)$ 
for several values of $\xi,\rho$, are shown at the bottom rows 
of Figs.~2-4.

The relevant potential in the spherical phase is 
$V_{1}(\rho;\beta,\gamma)$, Eq.~(\ref{eq:V1}), with 
$0\leq\rho\leq\rho_c$. 
In this case, $\beta_{\rm eq}=0$ is a global minimum of the potential 
at an energy $V_{\rm sph}=0$, representing the spherical equilibrium shape. 
The limiting value at the domain boundary is 
$V_{\rm lim} = V_{1}(\rho;\beta=\sqrt{2},\gamma) = 2h_{2}$.
For $\rho=0$, (the U(5) limit), the 
potential is independent of $\gamma$ and 
has $\beta_{\rm eq}=0$ as a single minimum. 
For $\rho>0$, the potential depends on $\gamma$, and 
$\beta=0$ remains a single minimum for $0\leq\rho<\rho^{*}$,  
At the spinodal point $\rho^{*}$, 
$V_{1}(\rho)$ acquires an additional deformed local minimum at an 
energy $V_{\rm def} >0$, and a barrier develops between the two minima.
The spherical and deformed minima cross and become degenerate at the 
critical point $(\rho_c,\xi_c)$. 

The relevant potential in the deformed phase is 
$V_{2}(\xi;\beta,\gamma)$, Eq.~(\ref{eq:V2}), with  $\xi\geq\xi_c$. 
In this case, $[\beta_{\rm eq}=\tfrac{2}{\sqrt{3}},\gamma_{\rm eq}=0]$ 
is a global minimum of the potential at an 
energy $V_{\rm def}=0$, representing the deformed equilibrium shape.
The limiting value of the domain boundary is
$V_{\rm lim} = V_{2}(\xi;\beta=\sqrt{2},\gamma) = (2+\xi)h_2$. 
The two potentials coincide at the critical point $(\rho_c,\xi_c)$, 
$V_2(\xi_c;\beta,\gamma)=V_1(\rho_c;\beta,\gamma)$, and the barrier height 
is $V_{\rm bar} \!=\! \frac{1}{2}h_2 (1-\sqrt{3})^2=0.268h_2$. 
The spherical minimum turns local in $V_{2}(\xi)$ for $\xi>\xi_c$ 
with energy $V_{\rm sph} = 4h_2\xi >0$, and disappears at the 
anti-spinodal point $\xi^{**}$. For $\xi > \xi^{**}$, $\beta=0$ 
turns into a maximum and the potential 
remains with a single deformed minimum, reaching the SU(3) limit 
at $\xi=1$.
\begin{figure}[!t]
\begin{center}
\includegraphics[width=0.99\linewidth]{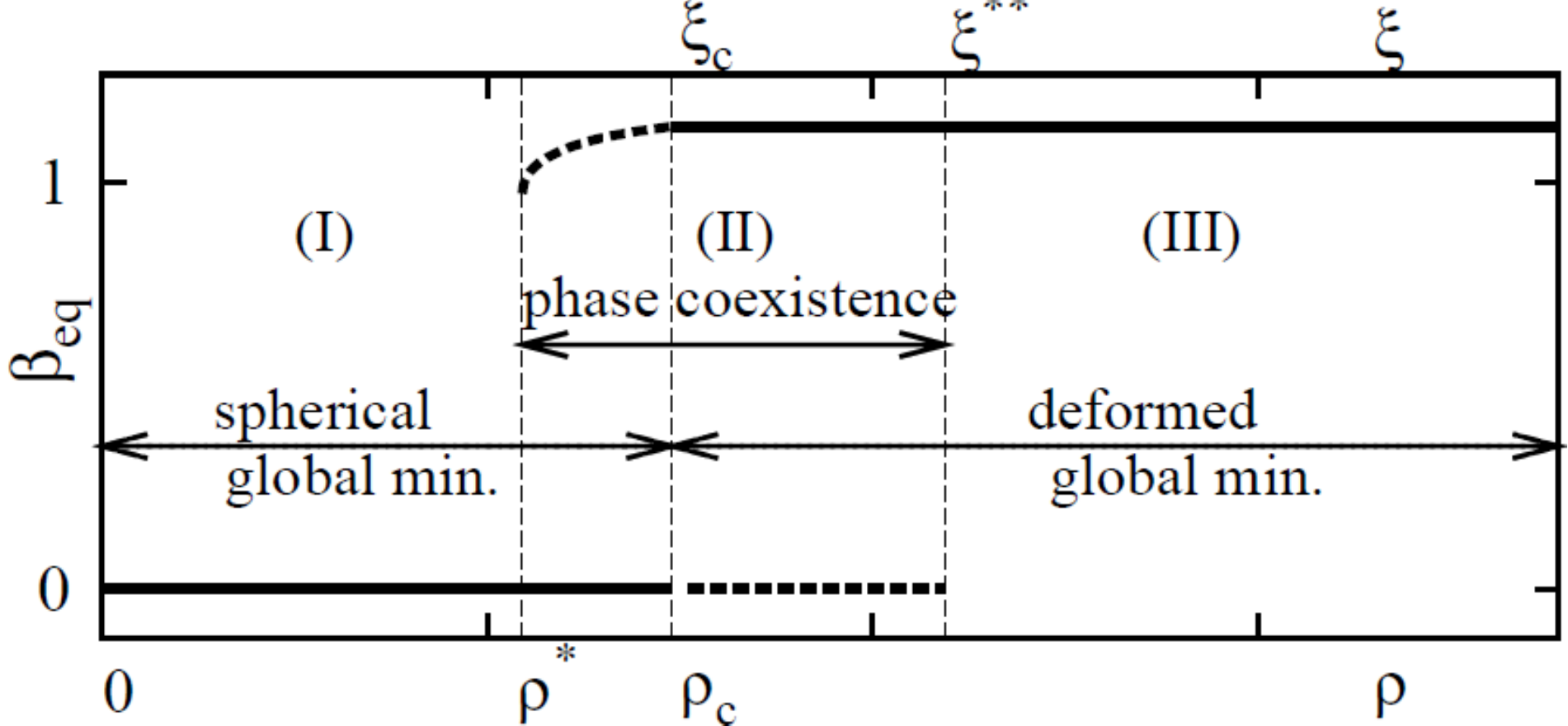}
\end{center}
\caption{Behavior of the order parameter, $\beta_{\mathrm{eq}}$, 
as a function of the control parameters ($\rho,\xi$) of the 
intrinsic Hamiltonian, Eq.~(\ref{eq:Hint}). Here 
$\rho^{*},\, (\rho_c,\,\xi_c),\, \xi^{**}$, are the spinodal, critical 
and anti-spinodal points, respectively, with values given in 
Eq.~(\ref{points}). The deformation at the 
global (local) minimum of the Landau potential (\ref{eq:Vcl}) is marked 
by solid (dashed) lines. 
$\beta_{\mathrm{eq}}\!=\!0$ ($\beta_{\mathrm{eq}}\!=\tfrac{2}{\sqrt{3}}$) 
on the spherical (deformed) side of the QPT.
Region I (III) involves a single spherical (deformed) shape, 
while region II involves shape-coexistence.}
\label{fig:1+-+-}
\end{figure}

The indicated changes in the topology of the potential surfaces 
upon variation of the control parameters $(\rho,\xi)$, identify 
three regions with distinct structure.
\begin{enumerate}
\item[I.]
The region of a stable spherical phase, $\rho \in [0,\rho^*]$, where 
the potential has a single spherical minimum.
\item[II.]
The region of phase coexistence, $\rho \in (\rho^*,\rho_c]$ and 
$\xi \in [\xi_c,\xi^{**})$, where 
the potential has both spherical and deformed minima which 
cross and become degenerate at the critical point. 
\item[III.]
The region of a stable deformed phase, $\xi > \xi^{**}$, where 
the potential has a single deformed minimum.
\end{enumerate}

The potential surface in each region serves as the Landau potential of the 
QPT, with the equilibrium deformations as order parameters. The latter 
evolve as a function of the control parameters ($\rho,\xi$) 
and exhibit a discontinuity typical of a first order transition. 
As depicted in Fig~1, 
the  order parameter ${\beta_{\mathrm eq}}$ is a 
double-valued function in the coexistence region (in-between $\rho^{*}$ 
and $\xi^{**}$) and a step-function outside it. In what follows, we examine 
the nature of the classical and quantum dynamics in each region.

\section{Classical analysis}

Chaotic properties of the IBM have been 
studied extensively~\cite{AW93}, albeit, with a simplified Hamiltonian, 
giving rise to an extremely low barrier. 
A new element in the present study is the presence of 
a high barrier at the critical-point, 
$V_{\rm bar}/h_2 \!=\!0.268$, compared to 
$V_{\rm bar}/h_2 \!=\!0.0018$ in previous works. 
This allows the uncovering of a rich pattern of regularity and chaos 
across a generic first-order QPT in a wide coexistence region.
\begin{figure*}[]
\begin{center}
\includegraphics[width=0.8\linewidth]{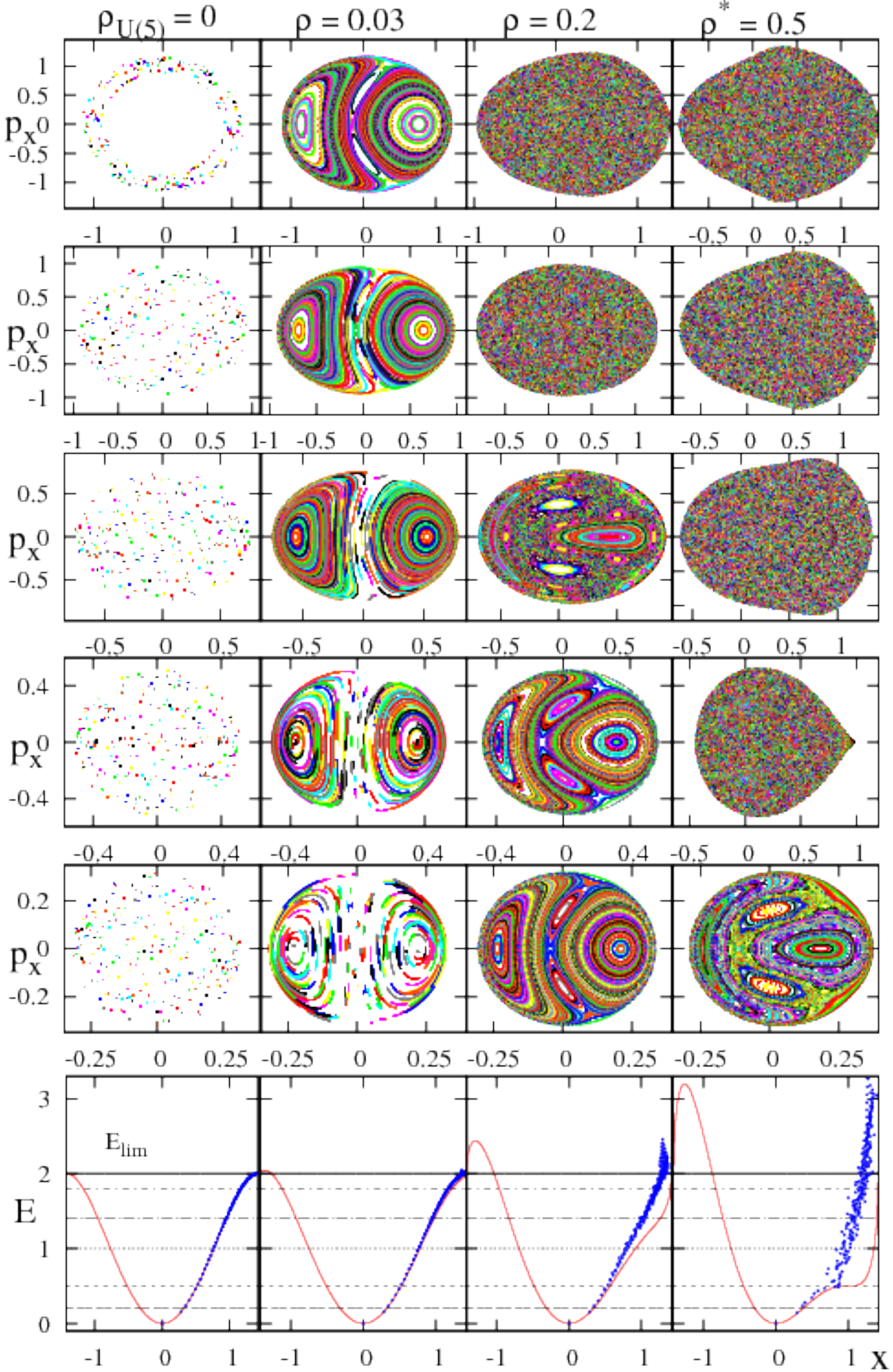}
\end{center}
\caption{
Poincar\'e sections in the stable spherical phase (region I). 
Upper five rows depict the classical dynamics of 
${\cal H}_{1}(\rho)$~(\ref{eq:H1cl}) 
with $h_2=1$, for several values of $\rho\leq\rho^{**}$. 
The bottom row displays the Peres lattices $\{x_i,E_i\}$, 
portraying the quantum dynamics for $(N=80,L=0)$ eigenstates 
of $\hat{H}_{1}(\rho)$~(\ref{eq:H1}), 
overlayed on the classical potentials 
$V_{1}(\rho;x,y=0)$~(\ref{eq:V1}). 
The five energies 
at which the sections were calculated consecutively, are indicated 
by horizontal lines.}
\end{figure*}
\begin{figure*}[]
\vspace{-1cm}
\begin{center}
\includegraphics[width=0.87\linewidth]{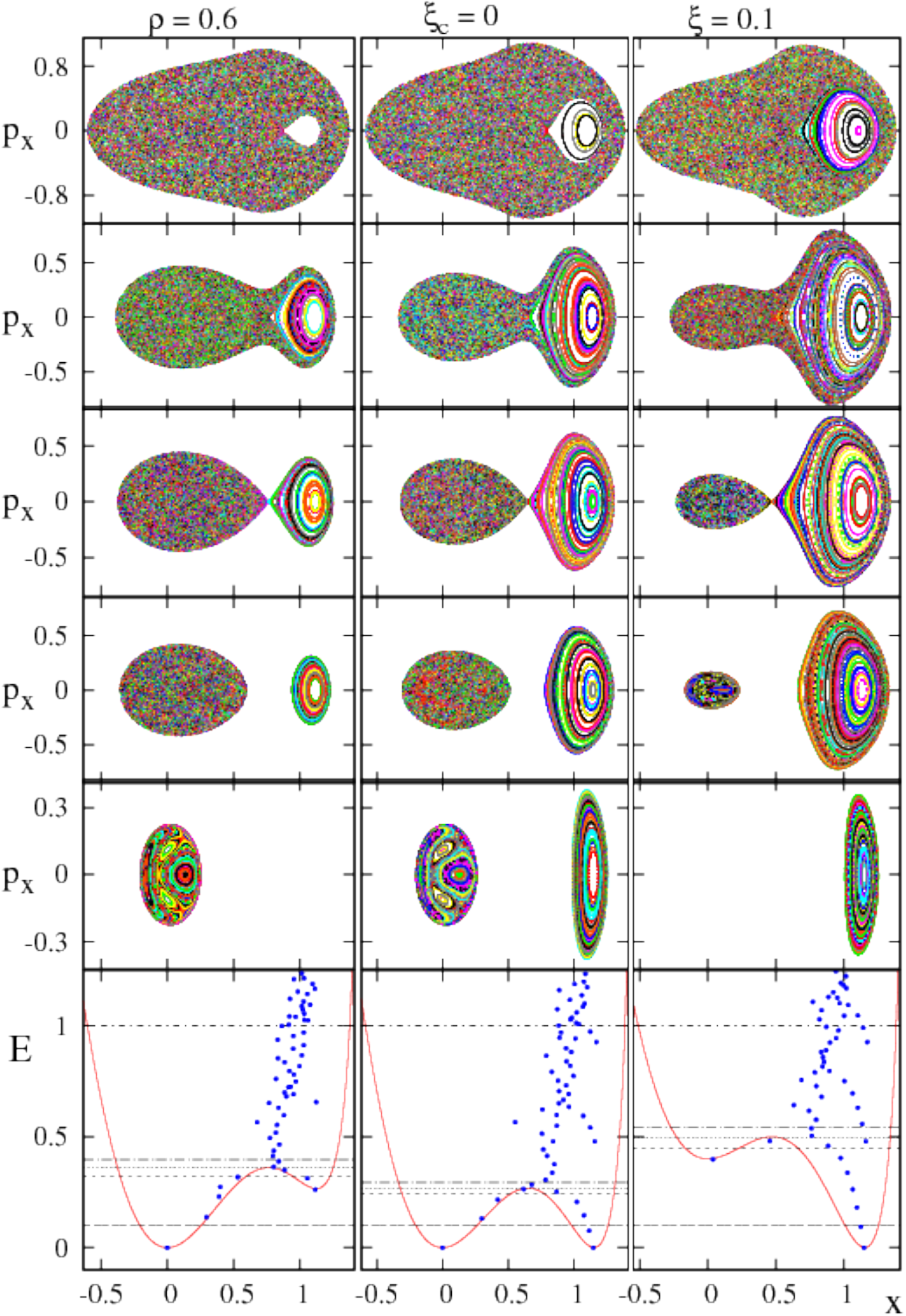}
\end{center}
\caption{
Poincar\'e sections in the region of phase-coexistence (region II). 
The panels are as in Fig.~2, but for the classical Hamiltonians 
${\cal H}_1(\rho)$~(\ref{eq:H1cl}) with $\rho^{**}<\rho \leq\rho_c$, 
and ${\cal H}_2(\xi)$~(\ref{eq:H2cl}) with $\xi_c\leq \xi < \xi^{**}$.
The classical potentials are $V_{1}(\rho;x,y=0)$~(\ref{eq:V1}) 
and $V_{2}(\xi;x,y=0)$~(\ref{eq:V2}). 
The Peres lattices involve the quantum Hamiltonians 
$\hat{H}_{1}(\rho)$~(\ref{eq:H1}) and $\hat{H}_{2}(\xi)$~(\ref{eq:H2}).}
\end{figure*}
\begin{figure*}[]
\vspace{-2cm}
\begin{center}
\includegraphics[width=0.9\linewidth]{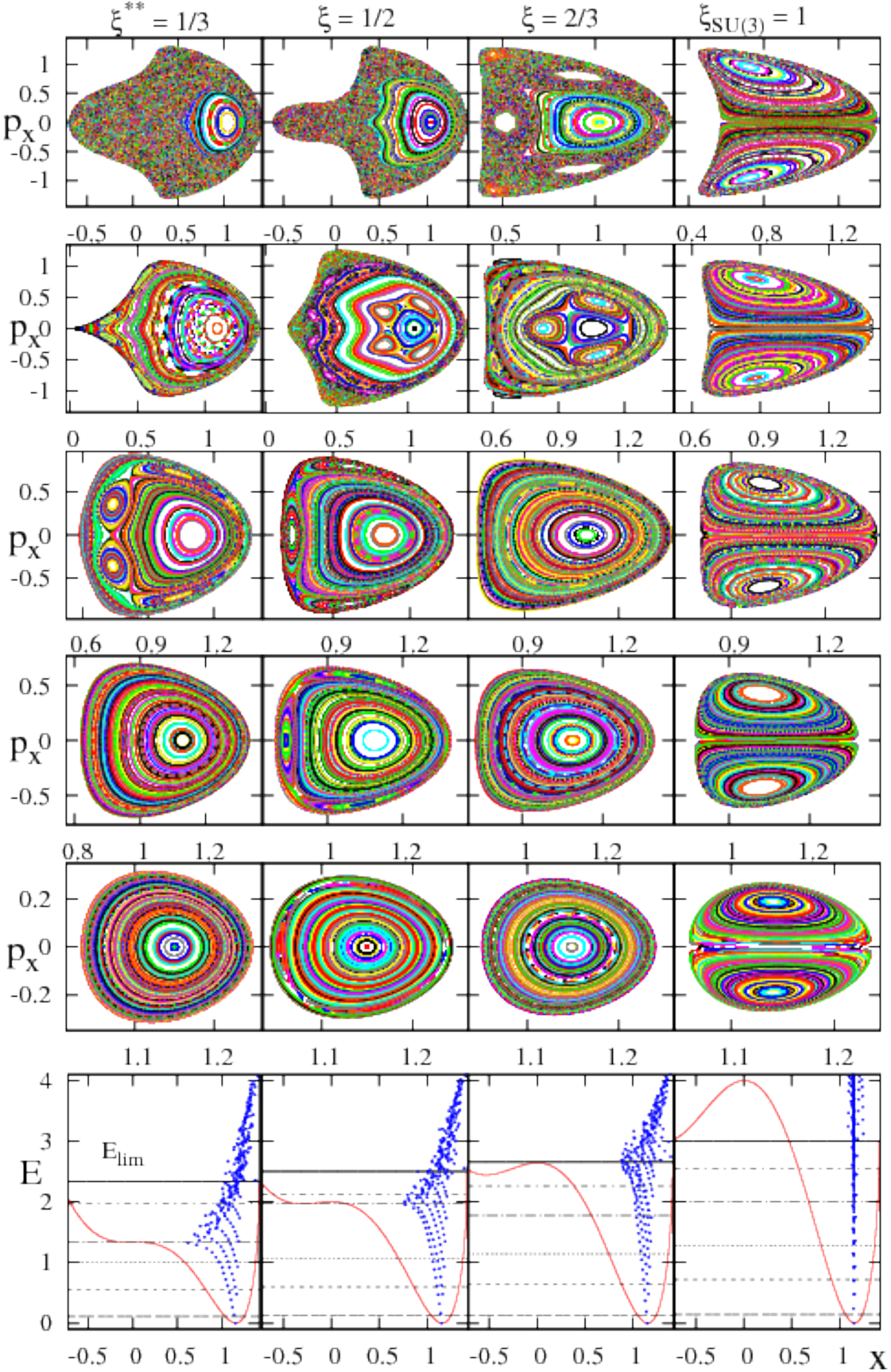} 
\end{center}
\caption{
Poincar\'e sections in the stable deformed phase (region III).
The panels are as in Fig.~2, but for the classical 
Hamiltonian ${\cal H}_2(\xi)$~(\ref{eq:H2cl}) and potential 
$V_{2}(\xi;x,y=0)$~(\ref{eq:V2}), with $\xi\geq\xi^{**}$. 
The Peres lattices involve the quantum Hamiltonian 
$\hat{H}_{2}(\xi)$~(\ref{eq:H2}).}
\end{figure*}

The classical dynamics constraint to $L=0$, 
can be depicted conveniently via 
Poincar\'e surfaces of sections in the plane $y=0$, 
plotting the values of $x$ and the momentum $p_x$ each time a 
trajectory intersects the plane~\cite{Reic92}. 
Regular trajectories are bound to toroidal manifolds within the 
phase space and their intersections with the plane of section lie 
on 1D curves (ovals). In~contrast, chaotic trajectories randomly 
cover kinematically accessible areas of the section.

The Poincar\'e sections associated with the classical 
Hamiltonian of Eq.~(\ref{eq:Hintcl})
are shown in Figs.~2-3-4 for 
representative energies, below the domain boundary, 
and control parameters ($\rho,\xi$) in regions~I-II-III, respectively. 
The bottom row in each figure 
displays the corresponding classical 
potential $V(\beta,\gamma\!=\!0)=V(x,y\!=\!0)$, Eq.~(\ref{eq:Vcl}).
In region~I $(0\leq\rho\leq\rho^{*}$), for $\rho=0$, 
${\cal H}_1(\rho\!=\!0)= {\cal H}_{d,0}(2-{\cal H}_{d,0})$, involves 
the 2D harmonic oscillator Hamiltonian and
$V_{1}(\rho\!=\!0)\!\propto\!2\beta^2 -\beta^4/2$.
The system is in the U(5) DS limit and is 
completely integrable. The orbits are periodic 
and, as shown in Fig.~2, appear in the surface of section 
as a finite collection of points. 
The sections for $\rho\!=\!0.03$ in Fig.~2, 
show the phase space 
portrait typical of an anharmonic (quartic) oscillator 
with two major regular islands, weakly perturbed by the 
small $\rho\cos 3\gamma$ term.
The orbits are quasi-periodic and appear as smooth one-dimensional 
invariant curves. 
For small~$\beta$, 
$V_{1}(\rho)\!\approx\! \beta^2 \!-\! 
\rho\sqrt{2}\beta^3\cos 3\gamma$ and resembles the 
well-known H\'enon-Heiles potential (HH)~\cite{Heno64}. 
Correspondingly, as shown for $\rho\!=\!0.2$ in Fig.~2, 
at low energy, the dynamics remains regular 
and two additional islands show up. 
The four major islands surround stable fixed points and 
unstable (hyperbolic) fixed points occur in-between.
At higher energy, one observes a marked onset 
of chaos and an ergodic domain.
The chaotic component of the dynamics increases with $\rho$ and 
maximizes at the spinodal point $\rho^{*}\!=\!0.5$.
The chaotic orbits densely fill two-dimensional regions of the 
surface of section.

The dynamics changes profoundly in region~II of phase coexistence 
$(\rho^{*} < \rho \leq\! \rho_c$ and $\xi_c \leq \xi < \xi^{**})$. 
The Poincar\'e sections 
before at and after the critical point, 
($\rho\!=\!0.6$, $\xi_c\!=\!0$, $\xi\!=\!0.1$) are shown in Fig.~3.
In general, the motion is predominantly regular at low energies and 
gradually turning chaotic as the energy increases. 
However, the classical dynamics evolves 
differently in the vicinity of the two wells. 
As the local deformed minimum develops, robustly regular dynamics 
attached to it appears. The trajectories form a single island 
and remain regular even 
at energies far exceeding the barrier height $V_{\rm bar}$.
This behavior is in marked contrast to the HH-type of dynamics 
in the vicinity of the spherical minimum, where a change with energy 
from regularity to chaos is observed, until complete chaoticity is reached 
near the barrier top. 
The clear separation between regular and chaotic dynamics, 
associated with the two minima, persists all the way to the 
barrier energy, $E=V_{\rm bar}$, where the two regions just touch. 
At $E > V_{\rm bar}$, the chaotic trajectories from the spherical region 
can penetrate into the deformed region 
and a layer of chaos develops, 
and gradually dominates the surviving regular island 
for $E\gg V_{\rm bar}$. As $\xi$ increases, the spherical minimum becomes 
shallower, and the HH-like dynamics diminishes.

Fig.~4 displays the classical dynamics in region~III 
$(\xi^{**}\leq\xi\leq 1)$. The local spherical minimum and the associated 
HH-like dynamics disappear at the anti-spinodal point $\xi^{**}=1/3$.
The regular motion, associated with the single deformed minimum, 
prevails for $\xi\!\geq\! \xi^{**}$. Here   
a single stable fixed point, surrounded by a family of 
elliptic orbits, continues to dominate the Poincar\'e section. 
In certain regions of the control parameter $\xi$ and energy, 
the section landscape changes from a single to several regular 
islands, reflecting the sensitivity of the dynamics 
to local degeneracies of normal modes. 
A notable exception to such variation is the SU(3) DS limit 
($\xi=1$), for which the system is integrable and the phase 
space portrait is the same for any energy. 

\section{Quantum analysis}

Quantum manifestations of classical chaos are 
often detected by statistical analyses of energy 
spectra~\cite{Reic92}. 
In a quantum system with mixed regular and irregular states, the 
statistical properties of the spectrum are usually intermediate between 
the Poisson and the Gaussian orthogonal ensemble (GOE) statistics. 
Such global measures of quantum chaos are, however, insufficient to 
reflect the rich dynamics of an inhomogeneous phase space structure  
encountered in Figs.~2-4, with mixed but well-separated regular and 
chaotic domains. 
To do so, one needs to distinguish between regular and irregular 
subsets of eigenstates in the same energy intervals. 
For that purpose, we employ the spectral lattice method of 
Peres~\cite{Peres84}, which 
provides additional properties of individual energy eigenstates. 
The Peres lattices are constructed by plotting the expectation 
values $O_i = \bra{i}\hat{O}\ket{i}$ of an arbitrary operator, 
$[\hat{O},\hat{H}]\neq 0$, versus the energy $E_i=\bra{i}\hat{H}\ket{i}$ 
of the Hamiltonian eigenstates $\ket{i}$. 
The lattices $\{O_i,E_i\}$ corresponding to 
regular dynamics can be shown to display an ordered pattern, while chaotic 
dynamics leads to disordered meshes of 
points~\cite{Peres84,Stran09}. 

In the present analysis we choose the Peres operator to be 
$\hat{n}_d$. The lattices correspond to the set of~points 
$\{x_i,E_i\}$, 
with $x_i \equiv \sqrt{2 \bra{i}\hat{n}_d\ket{i}/N}$ and $\ket{i}$ being 
the eigenstates of the quantum Hamiltonian~(\ref{eq:Hint}).
The expectation value of $\hat{n}_d$ in the condensate of 
Eq.~(\ref{condgen}) is related to the deformation $\beta$ 
and the coordinate $x$ in the classical potential~(\ref{eq:Vcl}). 
Accordingly, the particular choice of 
lattices $\{x_i,E_i\}$ can distinguish regular from irregular states 
and associate them with a given region in phase space, through the 
classical-quantum correspondence $\beta \!=\!x \!\leftrightarrow\! x_i$.

The Peres lattices for $L\!=\!0$ eigenstates of 
the intrinsic Hamiltonian~(\ref{eq:Hint}) 
with $N\!=\!80$, 
are shown on the bottom rows of Figs.~2-4, 
overlayed on the classical potentials $V(x,y=0)$ of Eq~(\ref{eq:Vcl}).
For $\rho\!=\!0$, the Hamiltonian~(\ref{H1u5}) 
has U(5) dynamical symmetry with a solvable spectrum, a function of $n_d$. 
For large $N$ and replacing $x_i$ by $\beta$, the Peres 
lattice coincides coincides with $V_{1}(\rho=0)$, 
a~trend seen exactly for $\rho=0$ and approximately at $\rho=0.03$ 
in Fig.~2. 
For $\rho=0.2$, at low energy a few lattice points still follow 
the potential curve $V_{1}(\rho)$, but at higher energies 
one observes sizeable deviations and disordered meshes of lattice points, 
in accord with the onset of chaos in the classical 
H\'enon-Heiles system. The disorder in the 
Peres lattice enhances at the spinodal point $\rho^{*}=0.5$, 
where the chaotic component of the classical dynamics maximizes.
As seen in Figs.~3-4, 
whenever a deformed minimum occurs in the potential, the Peres lattices 
exhibit regular sequences of states localized within and above the 
deformed well. They form several chains of lattice points 
close in energy, with the lowest chain 
originating at the deformed ground state. 
A close inspection reveals that the $x_i$-values 
of these regular states, lie in the intervals of $x$-values 
occupied by the regular tori in the Poincar\'e sections. 
Similarly to the classical tori, these regular 
sequences persist to energies well above the barrier $V_{\rm bar}$. 
The lowest sequence consists of 
$L\!=\!0$ bandhead states of 
the ground $g(K=0)$ and $\beta^n(K=0)$ bands. 
Regular sequences at higher energy correspond to 
$\beta^n\gamma^2(K=0)$, $\beta^n\gamma^{4}(K=0)$ 
bands, etc. 
In contrast, the remaining states, 
including those residing in the spherical minimum, 
do not show any obvious patterns and lead to 
disordered (chaotic) meshes of points at high energy. 
For $\xi>\xi^{**}$, 
a larger number and longer sequences of regular $K=0$ bandhead states 
are observed in the vicinity of the single deformed 
minimum ($x\approx 1$), as its depth increases. 
\begin{figure*}[!t]
\begin{center}
\includegraphics[width=0.9\linewidth]{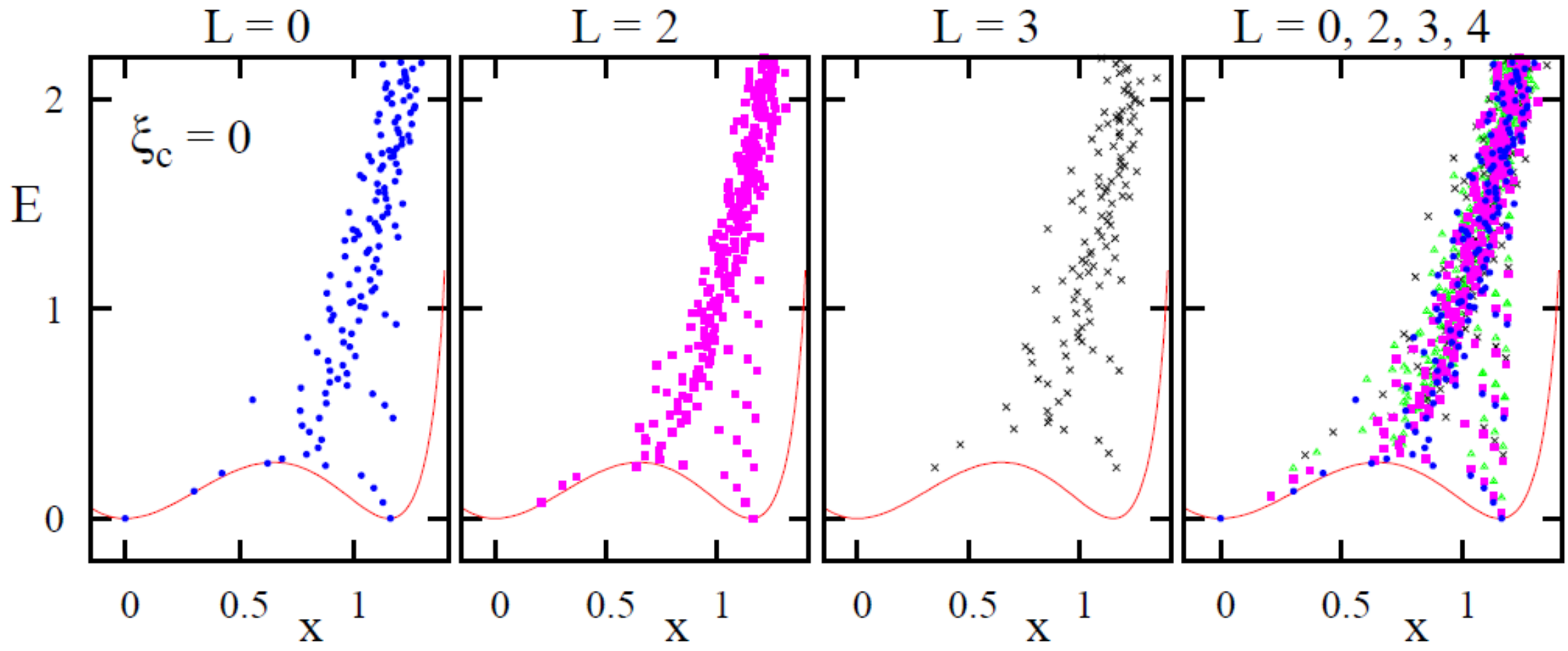}
\end{center}
\caption{Peres lattices $\{x_i,E_i\}$, 
for $L=0,2,3,4$, eigenstates 
of the critical-point Hamiltonian 
$\hat{H}_{1}(\rho_c)=\hat{H}_{2}(\xi_c)$ with $h_2=1$ and $N=50$. 
The right column combines the separate-$L$ lattices and overlays 
them on the corresponding classical potential.}
\end{figure*}

Peres lattices can also be used to visualize 
the dynamics of quantum states with non-zero angular momenta. 
Examples for $N\!=\!50$, $L=0,2,3,4$,
eigenstates of the critical-point Hamiltonian, 
$\hat{H}_{1}(\rho_c)=\hat{H}_{2}(\xi_c)$, are shown in Fig.~5.
The right column in the figure combines 
the separate-$L$ lattices and overlays them on the relevant 
classical potential. Rotational states with $L=0,2,4,\ldots$, 
comprise the regular $K\!=\!0$ bands mentioned above, 
and are accompanied by sequences $L=2,3,4,\ldots$, forming $K=2$ bands.
Additional $K$-bands (not shown in Fig.~5), 
corresponding to multiple $\beta$ and $\gamma$ vibrations 
about the deformed shape, can also be identified. 
The states in each regular band 
share a common intrinsic structure, as indicated by their nearly equal 
values of $\langle \hat{n}_d \rangle$ and a similar coherent decomposition 
of their wave functions in the SU(3) basis, to be discussed in Section~7.
These ordered band structures show up in the vicinity of the 
deformed well and are not present in the disordered (chaotic) portions 
of the Peres lattice. 
Their occurrence and persistence in the spectrum throughout 
the coexistence region, including the critical-point, 
is somewhat unexpected, in view of the strong mixing and abrupt 
structural changes taking place.

\section{Symmetry aspects}\label{sec:SymAsp}

Away from the U(5) and SU(3) limits ($\rho>0$ and $\xi<1$), 
both dynamical symmetries are broken in the 
intrinsic Hamiltonian~(\ref{eq:Hint}).  
The competition between terms with different symmetry character, 
drives the system through a first-order QPT with 
a characteristic pattern of mixed dynamics.
It is of great interest to study the symmetry properties of the 
Hamiltonian eigenstates across the QPT and, in particular, 
seek for a symmetry-based explanation for the persistence of regular subsets 
of states amidst a complicated environment.

Consider an eigenfunction of the Hamiltonian, $\ket{L_i}$, with 
angular momentum $L$ and ordinal number $i$ (enumerating the 
occurrences of states with the same $L$, with increasing energy). 
Its expansion in the U(5) DS basis, $\ket{N,n_d,\tau,n_{\Delta},L}$
of Eq.~(\ref{u5ds}), and in the SU(3) DS basis, 
$\ket{N,(\lambda,\mu),K,L}$ of Eq.~(\ref{su3ds}), reads
\ba
\ket{L_i}  &=& \sum_{n_d,\tau,n_{\Delta}}C^{(L_i)}_{n_d,\tau,n_{\Delta}}
\ket{N,n_d,\tau,n_{\Delta},L_i\,} ~, 
\nonumber\\
&=&
\sum_{(\lambda,\mu),K}C^{(L_i)}_{(\lambda,\mu),K}
\ket{N,(\lambda,\mu),K,L_i\,} ~.
\label{Li}
\ea
The U(5) ($n_d$) probability distribution, $P_{n_d}^{(L_i)}$, 
and the SU(3) [$(\lambda,\mu)$] probability distribution, 
$P_{(\lambda,\mu)}^{(L_i)}$, are calculated as
\bsub
\ba
P_{n_d}^{(L_i)} &=& \sum_{\tau,n_{\Delta}}
\vert C^{(L_i)}_{n_d,\tau,n_{\Delta}}\vert^2 ~,
\label{Pnd}\\
P_{(\lambda,\mu)}^{(L_i)} &=& \sum_{K}
\vert C^{(L_i)}_{(\lambda,\mu),K}\vert^2 ~.
\label{Plammu}
\ea
\label{Pndlammu}
\esub
The purity of eigenstates with respect to a DS basis
can be evaluated by means of the U(5) and SU(3) Shannon entropies
defined as
\bsub
\ba
S_\mathrm{U5}(L_i) &=& -\frac{1}{\ln D_{5}}
\sum_{n_d} P_{n_d}^{(L_i)} \ln P_{n_d}^{(L_i)} ~, 
\label{Shannonu5}\\
S_\mathrm{SU3}(L_i) &=& -\frac{1}{\ln D_{3}}
\sum_{(\lambda,\mu)} P_{(\lambda,\mu)}^{(L_i)} \ln P_{(\lambda,\mu)}^{(L_i)} ~. 
\label{Shannonsu3}
\ea
\label{eq:Shannon}
\esub
The normalization $D_5$ ($D_3$) counts the number of possible 
$n_d$ [$\lm$] values for a given $L$. 
A Shannon entropy vanishes when the considered state 
is pure with good $G$-symmetry
[$S_\mathrm{G}(L_i)\!=\!0$], and is positive for a mixed state. 
The maximal value [$S_\mathrm{G}(L_i)\!=\!1$] is obtained when 
$\ket{L_i}$ is uniformly spread among the irreps of $G$, 
{\it i.e.} for $P_{G}^{(L_i)}\!=\!1/D_G$. Intermediate values,  
$0 < S_\mathrm{G}(L_i) < 1$, indicate partial fragmentation 
of the state $\ket{L_i}$ in the respective DS basis.

\begin{figure}[!t]
\begin{center}
\includegraphics[width=0.493\linewidth]{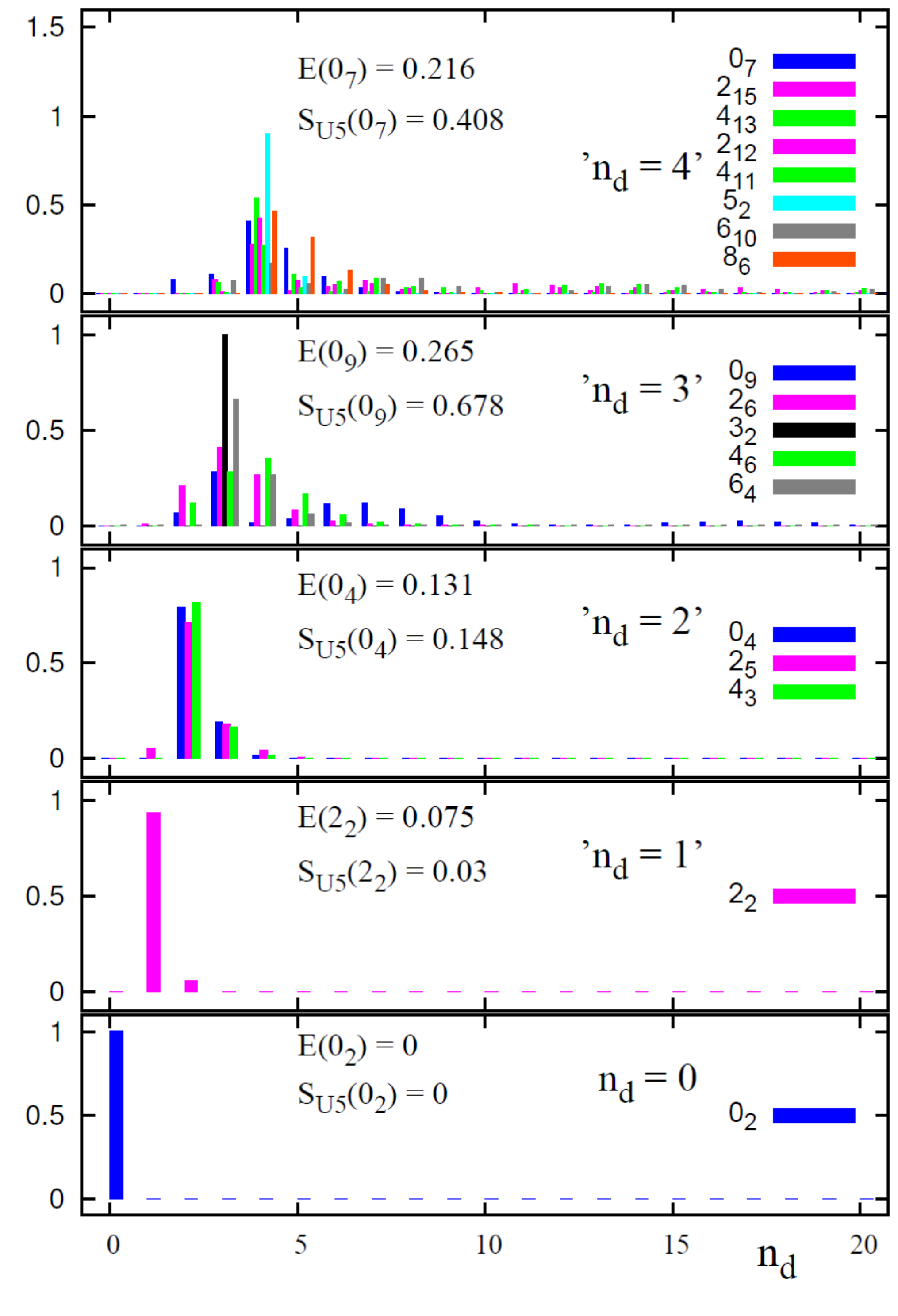}
\includegraphics[width=0.487\linewidth]{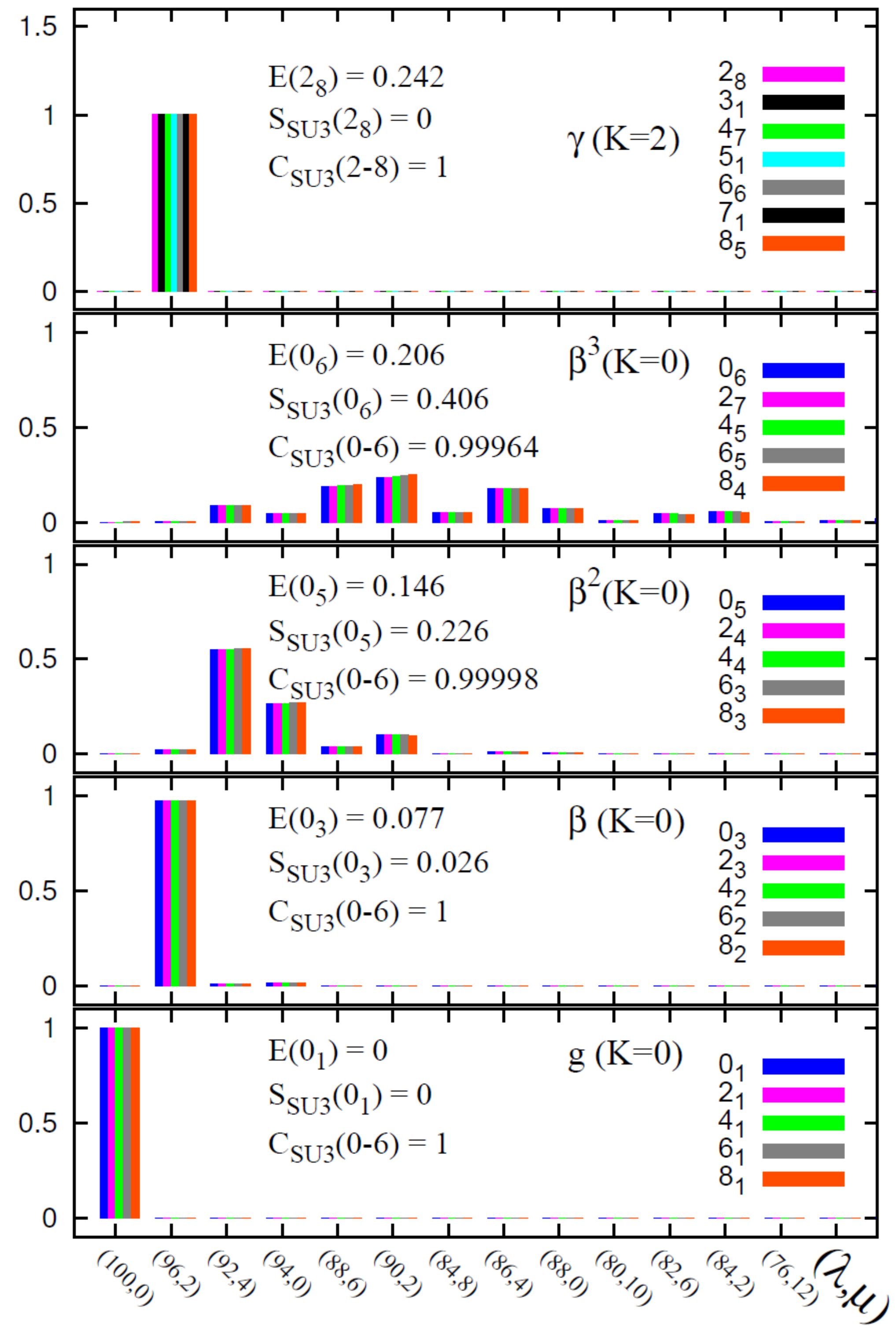}
\end{center}
\caption{U(5) $n_d$-probability distribution, $P_{n_d}^{(L_i)}$~(\ref{Pnd}) 
[left column], and SU(3) $(\lambda,\mu)$-probability distribution, 
$P_{(\lambda,\mu)}^{(L_i)}$~(\ref{Plammu}) [right column], for selected 
eigenstates of the critical-point $(\rho_c,\xi_c)$ 
Hamiltonian~(\ref{eq:Hint}), with $N=50$. 
The U(5) Shannon entropy, $S_{\rm U5}(L_i)$~(\ref{Shannonu5}),
and SU(3) correlator, $C_{\rm SU3}(0{\rm -}6)$~(\ref{Pearson}), 
are indicated for spherical and deformed type of states, respectively.}
\label{fig:7-1}
\end{figure}

Focusing on the critical-point Hamiltonian, 
the states shown on the left column of Fig.~6
were selected on the  
basis of having the largest components with $n_d=0,1,2,3,4$,
within the given $L$ spectra. States with different $L$ values 
are arranged into panels labeled by `$n_d$' to conform with the 
structure of the $n_d$-multiplets of the U(5) DS limit, Eq.~(\ref{H1u5}).
Each panel depicts the U(5) $n_d$-probabilities, 
$P_{n_d}^{(L_i)}$ (\ref{Pnd}), for states in the multiplet and lists 
the U(5) Shannon entropy $S_{\rm U5}(L_i)$ of a representative eigenstate.
In particular, the zero-energy $L\!=\!0^{+}_2$ state 
is seen to be a pure $n_d\!=\!0$ state, 
with $S_{\rm U5}\!=\!0$,  
which is the solvable U(5)-PDS eigenstate of Eq.~(\ref{ePDSu5L0}). 
The state $2^{+}_2$ has a pronounced $n_d\!=\!1$ 
component~(96\%) and the states ($L=0^{+}_4,\,2^{+}_5,\,4^{+}_3$) 
in the third panel, have a pronounced $n_d\!=\!2$ component and
a low value of $S_{\rm U5}< 0.15$.
All the above states with $`n_d\leq 2'$ have a dominant single $n_d$ 
component, and hence qualify as `spherical' type of states.
These multiplets comprise the lowest left-most states shown in the 
combined Peres lattices of Fig.~5. 
In contrast, the states in the panels `$n_d=3$' and `$n_d=4$' of 
Fig.~6, are significantly fragmented. Notable exceptions are 
the $L=3^{+}_2$ state, which is the solvable U(5)-PDS state of 
Eq.~(\ref{ePDSu5L3}) with $n_d=3$, and the 
$L=5^{+}_2$ state with a dominant $n_d=4$ component.
The existence in the spectrum of specific spherical-type of states with 
either $P_{n_d}^{(L)}\!=\!1$ $[S_{\rm U5}(L)\!=\!0]$ 
or $P_{n_d}^{(L)}\approx 1$ $[S_{\rm U5}(L)\approx 0]$, exemplifies 
the presence of an exact or approximate U(5) PDS at the critical-point.

The states shown on the right column of Fig.~6 have a different 
character. They belong to the five lowest regular sequences 
seen in the combined Peres lattices of Fig.~5. 
They have a broad $n_d$-distribution, hence are qualified as
`deformed'-type of states, forming rotational bands:
$g(K\!=\!0),\,\beta(K\!=\!0),\,\beta^2(K\!=\!0),
\,\beta^3(K\!=\!0)$ and $\gamma(K\!=\!2)$.
Each panel depicts the SU(3) $(\lambda,\mu)$-distribution, 
$P_{(\lambda,\mu)}^{(L_i)}$ (\ref{Plammu}) for the rotational states 
in each band. The ground $g(K=0)$ and the $\gamma(K=2)$  
bands are pure $[S_{\rm SU3}=0$] 
with $(\lambda,\mu) = (2N,0)$ and $(2N-4,2)$ SU3) character, respectively. 
These are the solvable bands of 
Eq.~(\ref{solsu3}) with SU(3) PDS. 
The non-solvable $K$-bands are mixed with respect 
to SU(3), but the mixing is similar for the different $L$-states 
in the same band. Such strong but coherent ($L$-independent) mixing is the 
hallmark of SU(3) quasi-dynamical symmetry (QDS). 
It results from the existence of a single 
intrinsic state for each such band and imprints an adiabatic motion and 
increased regularity~\cite{Dobes10}.

The coherent decomposition characterizing SU(3) QDS, implies 
strong correlations between the SU(3) components of different $L$-states 
in the same band. This can be used as a criteria for the identification 
of rotational bands. We focus here on the $L=0,2,4,6$, 
members of $K=0$ bands.
Given a $L=0^{+}_i$ state, among the ensemble of possible states, 
we associate with it those $L_j>0$ states which show the maximum 
correlation, $\max_{j}\{\pi(0_i,L_j)\}$.
Here $\pi(0_i,L_j)$ is a Pearson coefficient whose values lie
in the range $[-1,1]$.  Specifically, 
$\pi(0_i,L_j)=1,-1,0,$ indicate a perfect correlation, 
a perfect anti-correlation, and no linear correlation, respectively, 
among the SU(3) components of the $0_i$ and $L_j$ states. 
To quantify the amount of coherence (hence of SU(3)-QDS) in the chosen 
set of states, we employ the following product of the maximum correlation 
coefficients~\cite{MacDobCej10}
\ba
C_{\rm SU3}(0_i{\rm -}6) \equiv 
\max_{j}\{\pi(0_i,2_j)\}\,
\max_{k}\{\pi(0_i,4_k)\}\,
\max_{\ell}\{\pi(0_i,6_{\ell})\} ~.
\label{Pearson}
\ea
We consider the set of states $\{0_i,\,2_j,\,4_k,\,6_{\ell}\}$ as 
comprising a $K=0$ band with SU(3)-QDS, 
if $C_{\rm SU3}(0_i{\rm -}6)\approx 1$. 
As expected, we find the values $C_{\rm SU3}(0_i{\rm -}6)\approx 1$ for 
all the `deformed' $K$-bands, shown in the right column of Fig.~6. 
It should be noted that the coherence property of a band of states, 
as measured by $C_{\rm SU3}(0_i{\rm -}6)$, is independent of its purity, 
as measured by $S_{\rm SU3}(L_i)$. Thus, in Fig.~6, 
the pure $g(K=0)$ and $\gamma(K=2)$ bands with SU(3) PDS have 
$C_{\rm SU3}(0_i{\rm -}6)= 1$ and $S_{\rm SU3}=0$, 
while the mixed $\beta^3(K=0)$ band has 
$C_{\rm SU3}(0_i{\rm -}6)= 0.9996$ and $S_{\rm SU3}=0.406$.
\begin{figure}[!t]
\begin{center}
\includegraphics[width=0.92\linewidth]{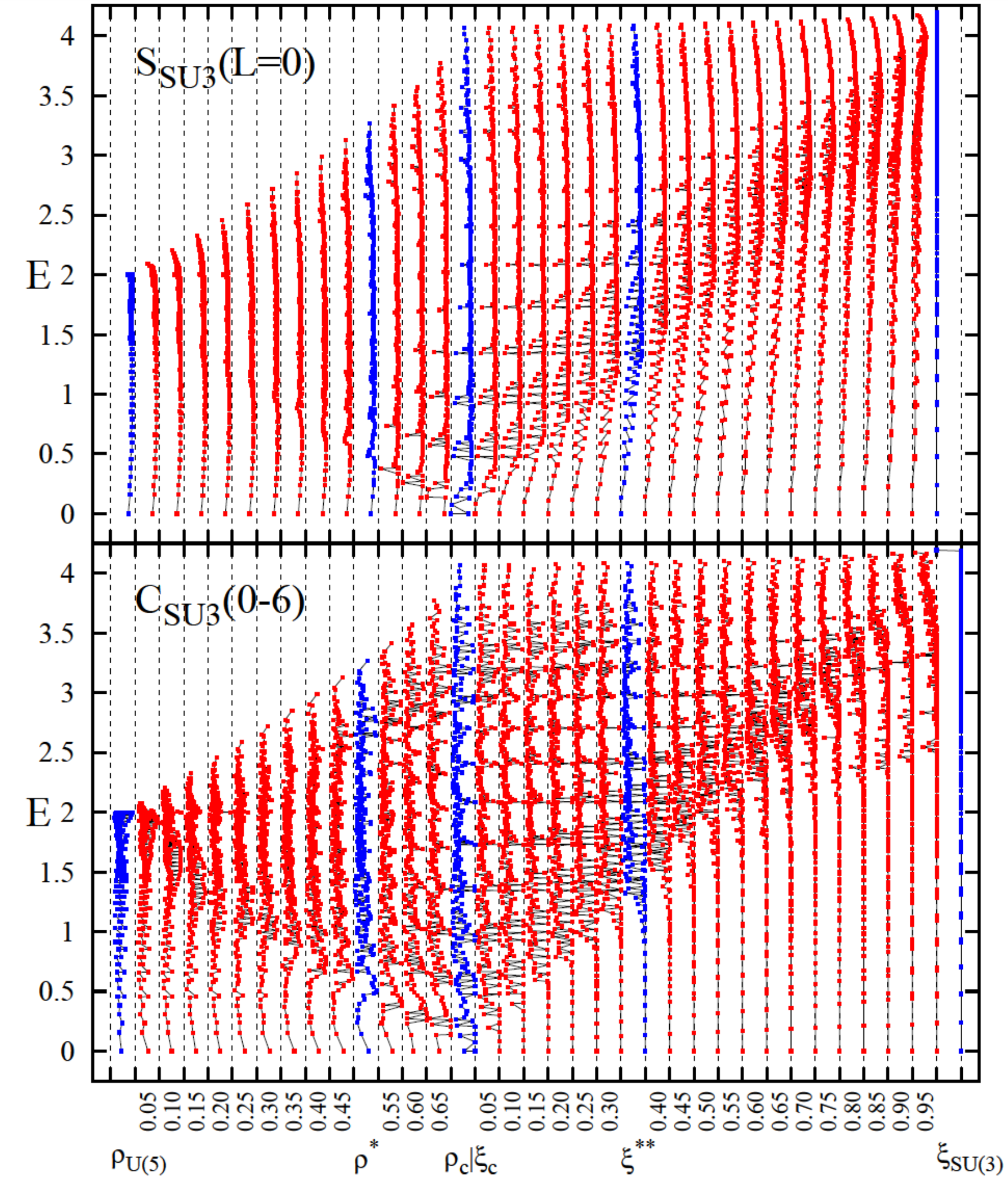}
\end{center}
\caption{SU(3) Shannon entropy $S_{\rm SU3}(L\!=\!0)$~(\ref{Shannonsu3}) 
[top panel], and SU(3) correlator, 
$C_{\rm SU3}(0{\rm -}6)$~(\ref{Pearson}) [bottom panel]
for energy eigenstates of the Hamiltonian~(\ref{eq:Hint}) 
with $N\!=\!50$, as a function of the control parameters $(\rho,\xi)$. 
The vertical dashed lines are explained in the~text.}
\label{fig:7-5}
\end{figure}

The top panel of Fig.~7 displays the values of the 
SU(3) Shannon entropy~(\ref{Shannonsu3}) 
for $L=0$ eigenstates of the Hamiltonian~(\ref{eq:Hint}), 
with $N=50$. The vertical axis lists the energy $E$ of the states, 
while the horizontal axis lists 35 values of the control parameters 
$(\rho,\xi)$. Vertical dashed lines which 
embrace each control parameter, correspond to the value 
$S_{SU3}(L\!=\!0)=0$ (left) and $S_{SU3}(L\!=\!0)=1$ (right). Thus, states which 
are pure with respect to SU(3) 
are represented by points on the 
vertical dashed line to the left of the given control parameter.
Departures from this vertical line, $0< S_{SU3}(L\!=\!0) \leq 1$, 
indicate the amount of SU(3) mixing. 
The bottom panel of Fig.~7, displays the values of the 
SU(3) correlation coefficient $C_\mathrm{SU3}(0{\rm -}6)$, 
Eq.~(\ref{Pearson}), correlating sequences of $L=0,2,4,6$ states, 
throughout the entire spectrum. 
The energy $E$, listed on the vertical axis, 
corresponds to the energy of the 
$L=0$ eigenstate in each sequence.
The vertical dashed lines 
correspond now to the value 
$C_\mathrm{SU3}(0{\rm -}6)=1$ (right) and 
$C_\mathrm{SU3}(0{\rm -}6)=0$ (left).
Thus, a highly-correlated sequence of $L=0,2,4,6$ states, comprising a 
$K=0$ band and manifesting SU(3)-QDS, 
are represented by points lying on or very close to the 
vertical dashed line to the right of the given control parameter, 
corresponding to $C_{\rm SU3}(0{\rm -}6)\approx 1$. 
Slight departures from this vertical line, 
$C_{\rm SU3}(0{\rm -}6)<1$, indicate a reduction of SU(3) coherence.

At the SU(3) DS limit ($\xi_{\rm SU(3)}=1$), the SU(3) entropy,
$S_{\rm SU3}(L) = 0$, vanishes for all states. 
In this case, the $L$-states in a given $K$-band belong 
to a single SU(3) irrep, hence necessarily $C_{\rm SU3}(0{\rm -}6)=1$. 
For $\xi<1$, 
$S_{\rm SU3}(L\!=0)>0$ acquires positive values, reflecting an SU(3) mixing. 
The SU(3) breaking becomes stronger at higher energies and 
as $\xi$ approaches $\xi_c=0$ from above, 
resulting in higher values of $S_{\rm SU3}(L\!=\!0)$. 
A notable exception to this behavior is 
the deformed ground state ($L=0_1$) 
of $\Htwo$, which maintains $S_{\rm SU3}(L\!=\!0_1) = 0$ throughout 
region~III ($\xi^{**}\leq\xi\leq 1$) and in part of 
region~II ($\xi_c\leq\xi<\xi^{**}$), in accord with 
its SU(3)-PDS property, Eq.~(\ref{solsu3g}). 
In contrast to the lack of SU(3)-purity in all excited 
$L=0$ states, the SU(3) correlation function maintains a value close 
to unity, $C_\mathrm{SU3}(0{\rm -}6)\approx 1$. This indicates that the SU(3) 
mixing is coherent and that these $L=0$ states serve as 
bandhead states of $K=0$ bands with a pronounced SU(3) QDS.
This band-structure is observed 
throughout region~III 
in extended energy domains, consistent 
with the classical analysis, which revealed a robustly 
regular dynamics in this region.

In region~I ($0\leq\rho\leq\rho^{*}$), all states 
show high values of $S_{\rm SU3}(L\!=\!0)\approx 1$ and 
$C_{\rm SU3}(0{\rm -}6)< 1$, indicating considerable SU(3)
mixing and lack of SU(3) coherence. This is in line with 
the presence of spherical-states, 
at low energy, and of more complex-type of states at higher energy, 
and the absence of rotational bands in this region.
In region~II of phase coexistence ($\rho^{*}<\rho\leq\rho_c$ and 
$\xi_c\leq\xi<\xi^{**}$), one encounters 
both points with $C_{\rm SU3}(0{\rm -}6)\approx 1$, and points with 
$C_{\rm SU3}(0{\rm -}6)< 1$. This reflects the presence of deformed states 
arranged into regular bands, exemplifying SU(3) QDS, and at the same time, 
the presence of spherical states and other 
states of a different nature. These results highlight the relevance 
of U(5)-PDS (partial purity) and SU(3)-QDS (coherence) for 
clarifying the survival of regular subsets of states in the presence of 
more complicated type of states, a situation encountered in QPTs of
non-integrable systems. 

\ack
This work is supported by the Israel Science Foundation. 
M.M. acknowledges the Golda Meir Fellowship Fund and the 
Czech Science Foundation (P203-13-07117S).

\end{document}